\documentclass[a4paper,conference]{IEEEtran}
\IEEEoverridecommandlockouts

\usepackage{cite}
\usepackage{amsmath,amssymb,amsfonts}
\usepackage{graphicx}
\usepackage{textcomp}
\usepackage[table]{xcolor}

\usepackage{orcidlink}
\usepackage{algorithm}
\usepackage{algpseudocode}
\usepackage{tabularx}
\usepackage{multirow}
\usepackage{url}
\usepackage{balance}
\usepackage{makecell}
\usepackage{eso-pic}

\newcommand{\IEEECopyrightNotice}{%
    \AddToShipoutPictureFG*{%
        \AtPageLowerLeft{%
            \put(\LenToUnit{0.65in},\LenToUnit{0.35in}){%
                \parbox{\dimexpr\paperwidth-1.3in\relax}{%
                    \fontsize{8}{9.5}\selectfont
                    \textcopyright{} 2026 IEEE. Personal use of this material is permitted. Permission from IEEE must be obtained for all other uses, in any current or future media, including reprinting/republishing this material for advertising or promotional purposes, creating new collective works, for resale or redistribution to servers or lists, or reuse of any copyrighted component of this work in other works.
                    \par
                }%
            }%
        }%
    }%
}
\hypersetup{hidelinks}

\def\BibTeX{{\rm B\kern-.05em{\sc i\kern-.025em b}\kern-.08em
    T\kern-.1667em\lower.7ex\hbox{E}\kern-.125emX}}
    
\begin{document}

\title{Confidential Model Sharing Protocol for Distributed\\Transfer Learning and Blockchain-Governed Access}

\author{Zeeshan Ali\textsuperscript{* \large \orcidlink{0000-0002-3577-7413}}, Walter Tiberti \textsuperscript{\large \orcidlink{0000-0002-4890-5031}}, Andrea Marotta \textsuperscript{\large \orcidlink{0000-0003-2426-1902}}, and Dajana Cassioli \textsuperscript{\large \orcidlink{0000-0001-7574-5099}}
\\
\IEEEauthorblockA{
Department of Information Engineering, Computer Science, and Mathematics (DISIM)\\
University of L'Aquila, L'Aquila, Italy\\
%Email: zeeshan.ali1@graduate.univaq.it, walter.tiberti@univaq.it, andrea.marotta@univaq.it, dajana.cassioli@univaq.it
}

\thanks{\textsuperscript{*}Corresponding author: Zeeshan ALI}

}

\maketitle
\IEEECopyrightNotice

\begin{abstract}
One of the main challenges of collaborative transfer learning (TL) is the \emph{secure} model sharing across distributed edge nodes. In this paper, we propose a novel protocol for the secure publication of models and controlled access in distributed environments. Any model artifact is locally encrypted and signed by the owner node, for integrity and authenticity verification, and then published to the InterPlanetary File System (IPFS) under a unique content identifier (CID). The CID and relevant model metadata are then registered immutably on an Ethereum smart contract via the publisher's blockchain wallet. The proposed protocol also introduces a blockchain-mediated access-request workflow in which a requesting node registers its public key on-chain, such that it can use its private key to locally recover the symmetric decryption key generated by the model owner, which records the access grant on-chain. The requester then verifies the publisher’s digital signature before decrypting. Our experimental evaluation demonstrates that the protocol supports transparent model lineage, selective access control, and end-to-end confidentiality, making it well-suited to secure TL in trustless distributed environments. We also provide considerations on the economic viability and energy efficiency of the proposed approach.
\end{abstract}


\section{Introduction}

\IEEEPARstart{I}{n} collaborative TL, participants shall exchange trained model artifacts without exposing intellectual property (IP) or relying on trusted centralized intermediaries \cite{pan2010survey,zeeshan2025deep}. Its widespread adoption across edge and distributed intelligence environments, however, has created a practical need for secure model exchange among mutually entrusted nodes \cite{kairouz2021advances}. 
In many real-world deployments, a node may train a \textit{basemodel} for a specific task and later share it with other participants for inference reuse, domain adaptation, or fine-tuning. However, direct model sharing exposes several security and governance problems. First, the model artifact itself may encode valuable IP and therefore cannot be released in plaintext. Second, a centralized repository introduces a single point of failure and weakens availability in geographically distributed settings. Third, traditional access control mechanisms provide limited transparency regarding ownership, provenance, model lineage, and permission updates. These limitations become more severe when multiple nodes iteratively refine shared models and when each participant needs cryptographic proof of authorship and integrity. To our knowledge, these issues have not yet been adequately addressed, preventing widespread use of TL for greater protection against cyber threats.
To improve trust and coordination in distributed learning environments, several works have explored blockchain-assisted ML systems. Early frameworks such as BlockFL demonstrated how blockchain can coordinate federated learning (FL) participants and provide incentive alignment in decentralized training environments \cite{kim2020blockfl}. Subsequent studies examined decentralized governance, smart-contract–mediated coordination, and the scalability and security implications of blockchain-enabled learning pipelines \cite{bhatia2022dfl}. While these approaches enhance auditability and participant accountability, they remain largely focused on coordinating training workflows rather than managing the secure publication, discovery, and reuse of trained model artifacts. 
Decentralized storage systems such as IPFS enable content-addressed distribution and resilient storage of large artifacts across peer-to-peer (P2P) networks \cite{benet2014ipfs}. Recent blockchain–IPFS integration studies have demonstrated practical architectures for secure storage and verifiable data access in decentralized environments \cite{alizouaghi2025ipfs}. 

\begin{figure*}[!htbp]
  \centering
  \includegraphics[width=0.73\linewidth]{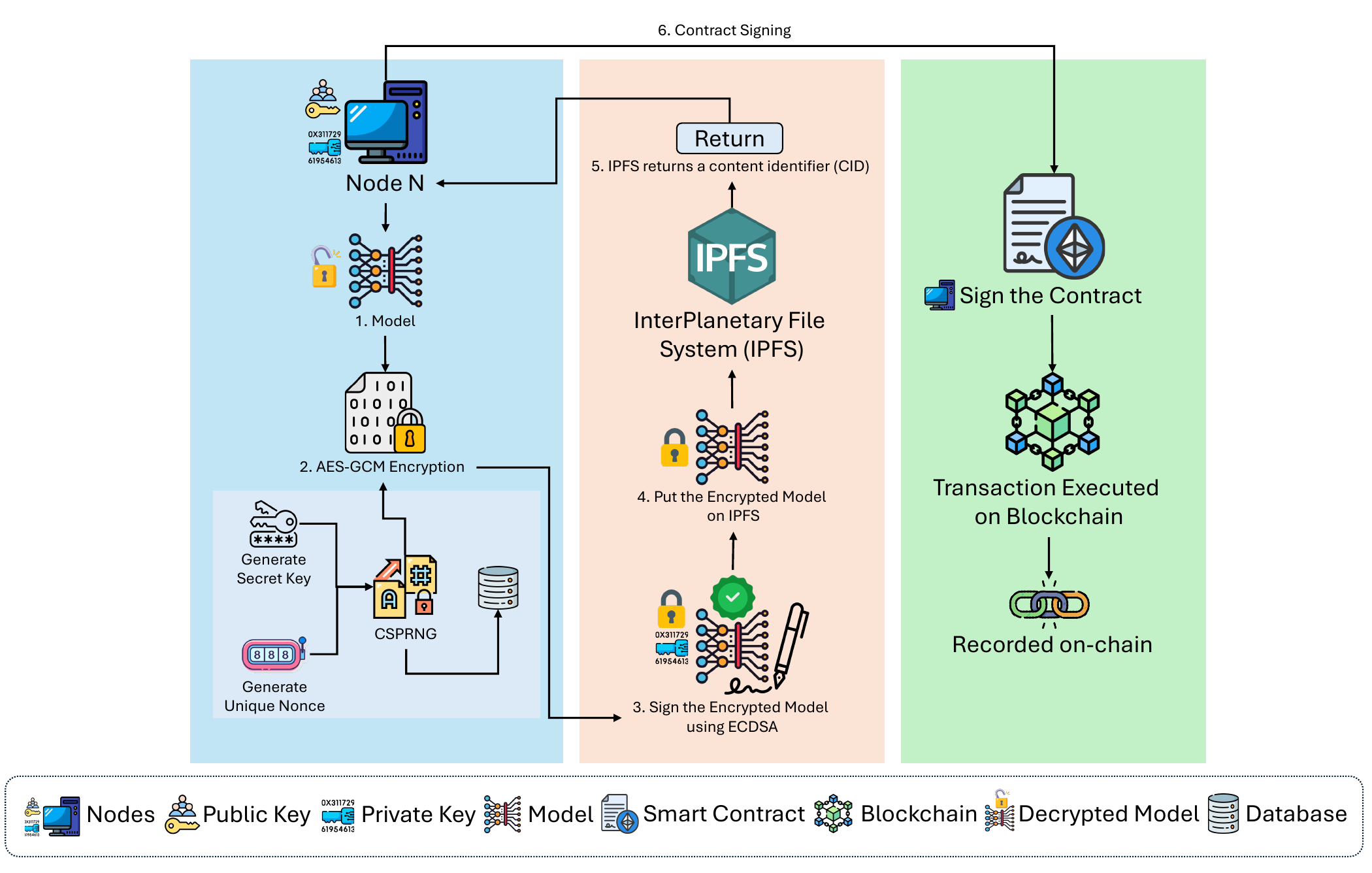}
  \caption{Protocol architecture for end-to-end secure TL, model encryption, IPFS integration, and blockchain-based provenance.}
  \label{Fig: System Model}
\end{figure*}

This work proposes a secure model-sharing protocol combining local cryptographic protection, IPFS-based distributed storage \cite{benet2014ipfs}, and blockchain-based access control \cite{wood2014ethereum}.
The proposed framework integrates three complementary layers for secure model dissemination in distributed environments. First, it separates data confidentiality from registry transparency by storing encrypted model artifacts in IPFS while recording searchable metadata and ownership evidence on-chain. Second, it employs a node-centric publication pipeline in which encryption, signature generation, and key retention occur locally before dissemination. Third, it introduces a blockchain-mediated access-request workflow combined with an off-chain ECDH-based key-wrapping mechanism \cite{haakegaard2015elliptic}, enabling model owners to selectively release the decryption key to approved requesters.

Consequently, models remain globally discoverable through their CID, while only authorized users can decrypt them. The proposed framework has been validated through experiments on a dedicated implementation of an end-to-end prototype supporting common model formats (e.g., ONNX and H5), encrypted IPFS publication, MetaMask-based transaction signing, and smart-contract-driven access management. The approach is demonstrated to be economically viable and energy efficient.

\section{Methodology}
\label{sec:methodology}

We consider a distributed TL environment composed of multiple autonomous edge nodes, as illustrated in Figure~\ref{Fig: System Model}. Node $N$ is an independent participant that may train a new \textit{basemodel}, publish an encrypted model artifact, discover previously shared models, request access to a protected model, or fine-tune an existing parent model in accordance with the policy defined by the owner. The design assumption is that nodes do not fully trust one another, yet they require a common infrastructure for verifiable publication and controlled reuse of model assets. Each node maintains the four logically distinct resources listed in Table~\ref{tab:node_resources}. This decomposition is intentional: computation and sensitive key material remain local to the node, encrypted artifacts are disseminated through decentralized storage, and only compact, verifiable metadata are committed to the blockchain.

Before publishing or requesting a model, each node $N_i$, for $i=1, 2, \ldots$, generates an asymmetric key pair $(pk_i, sk_i)$ using the Elliptic-Curve Cryptography (ECC). In this work we adopted the standard curve secp256k1, defined in the Standards for Efficient Cryptography (SEC) \cite{bib:secp}. The generated key pair is then encoded in Privacy-Enhanced Mail (PEM) format:

\begin{equation}
	(pk_i, sk_i) \leftarrow \mathsf{KeyGen}_{\text{secp256k1}}.
\end{equation}

The public key $pk_i$ identifies the node $N$ in the off-chain cryptographic layer.

\begin{table}[!bp] 
\centering 
\scriptsize
\caption{Node's Resources} 
\label{tab:node_resources} 
\begin{tabular}{|c|c|} 
\hline 
\rowcolor{gray!50} 
\textbf{Resource} & \textbf{Description} \\ \hline 
Local model repository & Plaintext model artifacts storage \\ 
\hline 
Local cryptographic state & \makecell{ECC key pair and symmetric keys for\\model encryption} \\ 
\hline 
IPFS distributed storage & Encrypted model envelopes publication \\ 
\hline 
Blockchain network & \makecell{A MetaMask wallet to authorize smart-\\contract transactions for model registration\\and blockchain access management} \\ 
\hline 
\end{tabular} 
\end{table}

When the node intends to publish a model, it creates a \emph{local model package} $\mathcal{P} = \{ M, \mathcal{A} \}$, where $M$ is the original model artifact and $\mathcal{A}$ is the associated descriptive metadata that can include architecture descriptor, preprocessing information, training context, version, and deployment-specific attributes.

Before publication, the node generates a fresh symmetric encryption key $K_M$ and a nonce $\eta$ using a secure pseudorandom number generator (CSPRNG):

\begin{equation}
	K_M \leftarrow \mathsf{CSPRNG}(256), \qquad \eta \leftarrow \mathsf{CSPRNG}(96).
\end{equation}
and encrypts the plaintext model artifact $M$ as $C_M = \mathsf{AES\text{-}GCM\_Enc}(K_M, \eta, M)$, where $C_M$ includes both the ciphertext and the authentication tag as produced by the Galois Counter Mode (GCM) mode.

\begin{figure*}[!htbp]
  \centering
  \includegraphics[width=0.60\linewidth, trim=0.21in 0.18in 0.29in 0.19in, clip]{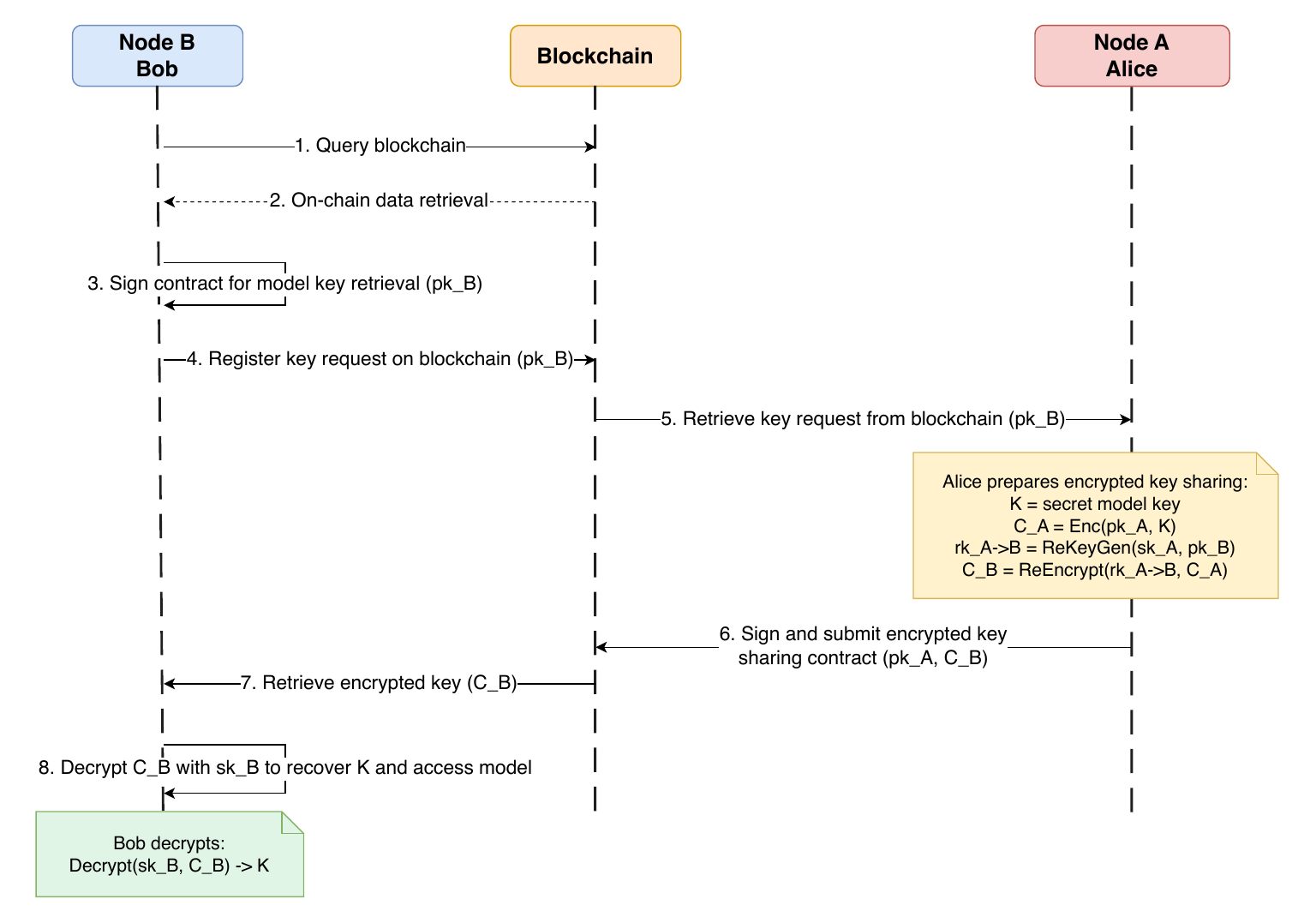}
  \caption{Sequence diagram of the access control layer to show the interaction between data requester (Bob), data owner (Alice), and the immutable ledger for key management.}
  \label{Fig: Model Discovery Sequence Diagram}
\end{figure*}

Notice that the key $K_M$ and the nonce are retained locally at the publisher node, hence, the IPFS receives only the encrypted model envelope and the decryption capability remains under the control of the originating node. To allow a requester to verify that the encrypted artifact was genuinely produced by the claimed publisher and has not been modified after publication, the publisher node computes a deterministic signature input $\Sigma_M$ on the encrypted model envelope using its private key, as

\begin{equation}
	\sigma_M = \mathsf{Sign}_{sk_i}(\mathcal{H}(\Sigma_M)),
\end{equation}

where $\mathcal{H}(\cdot)$ is instantiated with SHA-256 and the signing algorithm is Elliptic Curve Digital Signature Algorithm (ECDSA) over secp256k1. The $\sigma_M$ is stored in the local database alongside $K_M$, the model identifier, the original file name, and the algorithm descriptors, and embedded in the encrypted envelope that is uploaded to IPFS, together with the publisher's public verification key.

After encryption and signing, the node packages the protected artifact into an encrypted envelope, denoted by
\begin{equation}
	\mathcal{E}_M = \{ \texttt{modelId}, C_M, \sigma_M, pk_i, \texttt{metadata} \}.
\end{equation}

where \texttt{modelId} is a publisher-generated unique identifier for the model used to link the off-chain envelope to the on-chain registry record (i.e., $id_M$), and \texttt{metadata} is a structured descriptor that includes publication-time parameters as discussed in Table~\ref{tab:blockchain-record}, later registered on-chain.

The envelope is serialized as a \texttt{.enc.json} object and uploaded to IPFS, where the content is addressed by a hash-derived unique identifier $\mathsf{CID}(\mathcal{E}_M)$ and not by a server location. The CID is a cryptographic pointer to the encrypted envelope described as $\mathsf{CID}_M \leftarrow \mathsf{IPFS\_Add}(\mathcal{E}_M)$, that becomes the reference used by the blockchain registry. The CID does not reveal the plaintext model parameters; it merely provides a globally resolvable handle to the encrypted content.
The usage of IPFS has the advantage that, unlike centralized object storage, the dependency on a single administrative endpoint is reduced and resilience in distributed deployments is significantly improved.

\subsection{Blockchain Registration and Global Discoverability} %of Model Metadata}

At this point, the publisher node registers the model on the blockchain, 
by providing a record for model $M$ given by:
\begin{equation}
\begin{aligned}
R_M = \{ &id_M, \mathsf{CID}_M, owner_M, t_M, task_M, domain_M \\
         & ver_M, policy_M, parent_M \}
\end{aligned}
\end{equation}
where the descriptive metadata and notations are presented in Table~\ref{tab:blockchain-record}.

\begin{table}[!t]
\centering
\scriptsize
\caption{On-chain Model Record $R_M$ Description}
\label{tab:blockchain-record}
\footnotesize
\setlength{\tabcolsep}{3pt}

\begin{tabularx}{\columnwidth}{|c|X|}
\hline
\rowcolor{gray!50}
\textbf{Field} & \textbf{Description} \\\hline
$id_M$ & Unique identifier of the model \\\hline
$\mathsf{CID}_M$ & Content Identifier (CID) of the model stored in IPFS \\\hline
$owner_M$ & Blockchain address of the model owner \\\hline
$t_M$ & Model type (e.g., CNN, Transformer, ResNet, LSTM, GAN) \\\hline
$task_M$ & Task type (e.g., Classification, Detection, Segmentation) \\\hline
$domain_M$ & Application domain (e.g., IIoT, Healthcare, AVs, IoT) \\\hline
$ver_M$ & Version of the model \\\hline
$policy_M$ & Access policy (e.g., Inference-only, TL allowed) \\\hline
$parent_M$ & Parent model identifier (for lineage tracking) \\
\hline
\end{tabularx}
\end{table}

The smart contract emits corresponding events when a model is deployed or when a fine-tuned descendant is registered.

Thereafter, the model becomes discoverable to other participants through blockchain queries over attributes such as task type, domain, model type, owner, version, or access policy. However, the retrieved object remains encrypted and cannot be decrypted without access to $K_M$. This distinction between discoverability and decryptability is central to the proposed design.

\begin{table*}[htbp] 
\centering 
\scriptsize
\caption{Measured latency breakdown over 30 randomized model runs.} \label{tab:latency_breakdown} 
\begin{tabular}{|c|c|c|c|} \hline 
\rowcolor{gray!50}
\textbf{Operation} & \textbf{Median (ms)} & \textbf{P95 (ms)} & \textbf{Notes} \\\hline 
Key generation (secp256k1) & 0.746 & 1.430 & Per requester keypair generation \\\hline
AES-256-GCM encryption & 44.738 & 126.125 & Strongly model-size dependent \\\hline 
ECDSA signing & 89.473 & 188.457 & Payload-envelope integrity \\\hline 
IPFS upload + CID return & 329.813 & 876.116 & Network and local IPFS daemon dependent \\\hline 
ECDH key wrapping & 1.474 & 2.604 & Per access grant \\\hline
ECDH key unwrapping & 1.337 & 3.282 & Requester-side recovery \\\hline 
AES-256-GCM decryption & 34.980 & 114.357 & Decryption-side model recovery \\\hline  
\end{tabular} 
\end{table*}

\begin{table*}[!htbp] 
\centering 
\scriptsize
\caption{Measured on-chain cost by action type across Sepolia and Polygon Amoy (POL) deployments.} 
\label{tab:fee_breakdown} 
\begin{tabular}{|c|c|c|c|c|c|} 
\rowcolor{gray!50}
\hline 
\textbf{Network} & \textbf{Action} & \textbf{Avg. Gas Used} & \textbf{Avg. Fee (Native)} & {USD (Reference)} & \textbf{Main Cost Driver} \\ \hline 
Sepolia & Model registration ($n=41$) & 347{,}795 & 0.000701313 SepoliaETH & 1.639539 & Metadata string writes \\\hline  
Sepolia & Access request ($n=11$) & 328{,}858 & 0.000493286 SepoliaETH & 1.153212 & Public-key string storage \\\hline  
Sepolia & Access grant ($n=5$) & 635{,}524 & 0.000953286 SepoliaETH & 2.228606 & Encrypted-key payload storage \\\hline  
POL & Model registration ($n=11$) & 342{,}260 & 0.008898765 POL & 0.000889 & Metadata string writes \\\hline  
POL & Access request ($n=10$) & 327{,}665 & 0.008519290 POL & 0.000851 & Public-key string storage \\\hline  
POL & Access grant ($n=10$) & 632{,}142 & 0.016435692 POL & 0.001642 & Encrypted-key payload storage \\ \hline 
\end{tabular} 
\end{table*}

\subsection{Blockchain-Mediated Model Access}

To obtain a model for inference or fine-tuning, a requester node (e.g., Bob) first identifies a suitable model by querying the blockchain registry, as shown in the sequence diagram of Figure~\ref{Fig: Model Discovery Sequence Diagram}. The search criteria may include a target task, application domain, or an access policy indicating whether downstream TL is permitted. Suppose Node B (Bob) wishes to access a model published by Node A (e.g., Alice). Let the selected model identifier be $id_M$, and let Node B's public key be $pk_B$.
Node B submits an access request to the smart contract by providing the model identifier $id_M$ and its public key $pk_B$, along with its blockchain identity. The request is recorded on-chain as $Q_{B \rightarrow A} = \{ id_M, addr_B, pk_B, t_s \}$ and includes the requester’s wallet address $addr_B$, public key $pk_B$, and timestamp $t_s$.
After the transaction is confirmed, the model owner can view and manage the pending request through the blockchain-based request management interface.
Upon receiving a valid request, the model owner Node A (Alice) evaluates the received key request. If access is approved, the owner performs node-local ECDH-based wrapping of the locally stored model key $K_M$ using the requester's public key $pk_B$. This is a key-wrapping operation rather than true PRE, because the owner computes the transformation locally and no proxy re-encrypts an existing ciphertext. The owner then generates an ephemeral secret and derives a shared wrapping key, after which the model key is protected for the requester as $\widetilde{K}_{M \rightarrow B} = \mathsf{ECDH\_Wrap}(K_M, pk_B)$.

The resulting protected key object contains the ephemeral public key, initialization vector, authentication tag, and ciphertext. This object is then submitted to the access-grant smart contract together with the model identifier, the requester's wallet address, and the requester's public key. The blockchain, therefore, stores an auditable grant record while the sensitive transformation itself remains off-chain.
Once the key grant is published, Node B (Bob) retrieves the protected key object from the blockchain-accessible grant record. Using its own private key $sk_B$, Node B derives the corresponding shared secret, recovers the wrapped model key, and obtains the original symmetric decryption key by $K_M = \mathsf{ECDH\_Unwrap}(\widetilde{K}_{M \rightarrow B}, sk_B)$.
Node B then downloads the encrypted model envelope from IPFS using $\mathsf{CID}_M$, verifies the embedded ECDSA signature using the publisher's public key $pk_A$, and decrypts the encrypted model payload using AES-256-GCM to obtain $M = \mathsf{AES\text{-}GCM\_Dec}(K_M, \eta, C_M)$.
If the model's access policy permits inference only, Node B uses the decrypted model in a read-only operational setting. If fine-tuning is allowed, Node B can adapt the model to local data and publish the resulting descendant model as a new registry entry with the original model referenced in the field $parent_M$. In this way, the blockchain captures model lineage across successive TL stages.
This lineage structure enables at least three useful capabilities: transparent provenance analysis, traceability of collaborative TL, and policy-aware reuse of downstream models.

The proposed architecture satisfies the key requirements of secure model sharing in distributed TL: model confidentiality is enforced through node-local symmetric encryption, integrity and authenticity are provided through digital signatures, availability is improved through distributed IPFS storage, and ownership, discovery, access control, and lineage are made transparent through blockchain registration.

\section{Results and Discussion}
\label{sec: results} 

This section evaluates the performance of the proposed secure TL pipeline with emphasis on execution latency and deployment feasibility. The implemented system integrates local cryptographic processing (AES-256-GCM, ECDSA-secp256k1, and ECDH-based key wrapping), decentralized artifact storage using IPFS, and smart-contract execution for model registration and access control.

Performance measurements were collected using an automated benchmark over $n=30$ randomized runs with model artifacts between 20 MB and 100 MB. Each run executed the full processing pipeline used in the system: secp256k1 key generation, AES-256-GCM encryption, ECDSA-secp256k1 signing, ECDH-based key wrapping/unwrapping, AES-256-GCM decryption, and IPFS upload latency measurement.
The generated dataset ranged from 24,129,721 bytes to 99,536,480 bytes, with a mean size of 60,724,332 bytes ($\approx57.9\,\text{MB})$. To make the variance explicit while keeping the table compact, we report both median and 95th percentile (P95) latencies in Table~\ref{tab:latency_breakdown}; the randomized run design and fixed artifact-size range also support reproducibility.
Measurements show that cryptographic key-management operations introduce minimal overhead. Key generation and ECDH-based key wrapping/unwrapping complete within approximately ($\approx1-3\,\text{ms})$, even at the P95 level. In contrast, payload-dependent operations dominate the runtime. AES-256-GCM encryption and decryption scale with artifact size, while ECDSA signing reflects the cost of generating integrity proofs for the encrypted model envelope.
The most significant contributor to latency is the IPFS publication stage. As shown in Table~\ref{tab:latency_breakdown}, the IPFS upload and CID generation process reaches a median latency of ($\approx329\,\text{ms})$ and a P95 of ($\approx876\,\text{ms})$, reflecting network and local daemon variability. This confirms that storage and network transfer, rather than cryptographic processing, dominate the end-to-end pipeline delay. Importantly, the ECDH-based key wrapping operations remain stable and low-latency, supporting efficient access-grant workflows even for large model artifacts.
To evaluate deployment behavior across blockchain environments, the protocol was executed on two Ethereum-compatible test networks: Sepolia and POL. This multi-network validation ensures that smart-contract execution, access-control workflows, and metadata registration behave consistently across different fee markets.
Table~\ref{tab:fee_breakdown} reports the observed gas consumption and transaction costs for the primary protocol operations: model registration, access request submission, and access-grant authorization. For each action type, we collected the corresponding transactions generated during the experiments (sample size $n$ in Table~\ref{tab:fee_breakdown}) and computed the averages from their on-chain receipts using a Python script. For each transaction, we read \texttt{gasUsed} and \texttt{effectiveGasPrice}, compute the native fee as $\mathsf{fee}=\texttt{gasUsed}\times\texttt{effectiveGasPrice}$, and then report the mean \texttt{gasUsed} and mean $\mathsf{fee}$ over the $n$ transactions. Gas usage remains largely consistent across networks because it is determined by the computational and storage complexity of the smart-contract logic, whereas the resulting monetary cost varies due to network-specific fee markets and token evaluations.
In the experiments, POL produced significantly lower USD-denominated transaction costs than Sepolia while maintaining identical protocol behavior. The primary contributors to gas consumption are metadata string storage during model registration, requester public-key storage during access requests, and encrypted key-payload storage during access grants.
For completeness, the values reported in Table~\ref{tab:fee_breakdown} were obtained from validation-time observations on test networks and should be interpreted as comparative engineering indicators rather than fixed production costs. Since testnet native tokens (Sepolia and POL) are not directly priced, the USD column is provided only as a reference conversion by applying contemporaneous mainnet token/USD spot valuations.

These findings demonstrate that while the protocol is functionally portable across Ethereum-compatible networks, economic efficiency is network-dependent. Consequently, practical deployment should treat blockchain selection and fee-policy tuning as part of overall system optimization alongside cryptographic workflow and storage design.
The results depend on the current implementation, workload characteristics, and evaluation environment. Network-dependent measurements obtained on Sepolia and POL represent validation-time observations rather than fixed production deployment costs. To mitigate this limitation, the study reports operation-level timing, provides reproducible benchmarking scripts, and treats testnet cost figures as comparative indicators rather than deterministic production estimates.

\section{Conclusion and Future Work}
\label{sec: conclusion}

This work presents a secure protocol for distributed model sharing that integrates node-local cryptography, IPFS-based distributed storage, and blockchain-based metadata governance. The framework also supports blockchain-mediated access requests and owner-authorized key sharing through an ECDH-based key-wrapping mechanism, allowing authorized nodes to obtain decryption capability without exposing publisher secrets or plaintext models on-chain. Experimental evaluation shows that cryptographic key-management operations introduce minimal latency, while payload size and IPFS publication dominate end-to-end delay. Cross-network experiments on Sepolia and POL demonstrate that the protocol is portable across Ethereum-compatible networks, although economic behavior varies by network. Future work will incorporate a mandatory pre-publication model analysis layer to mitigate malicious model dissemination.

\section*{Acknowledgement}
This work is partially supported by the European Commission through the HORIZON-MSCA-SE EXARCH project (GA: 101236244).

\bibliography{bibliography}

@article{zeeshan2025deep,
  title={Deep Transfer Learning for Intrusion Detection in Edge Computing Scenarios},
  author={Zeeshan, ALI and Tiberti, Walter and Marotta, Andrea and Cassioli, Dajana},
  journal={IEEE Internet of Things Journal},
  year={2025},
  publisher={IEEE}
}

@article{kairouz2021advances,
	author  = {Peter Kairouz and H. Brendan McMahan and Brendan Avent and others},
	title   = {Advances and Open Problems in Federated Learning},
	journal = {Foundations and Trends in Machine Learning},
	volume  = {14},
	number  = {1--2},
	pages   = {1--210},
	year    = {2021}
}

@article{kim2020blockfl,
	author  = {Hyesung Kim and Jihong Park and Mehdi Bennis and Seong-Lyun Kim},
	title   = {Blockchained On-Device Federated Learning},
	journal = {IEEE Communications Letters},
	volume  = {24},
	number  = {6},
	pages   = {1279--1283},
	year    = {2020}
}

@misc{wood2014ethereum,
	author       = {Gavin Wood},
	title        = {Ethereum: A Secure Decentralised Generalised Transaction Ledger},
	howpublished = {Ethereum Yellow Paper},
	year         = {2014},
	note         = {Accessed: 2026-03-16}
}

@misc{benet2014ipfs,
	author       = {Juan Benet},
	title        = {IPFS - Content Addressed, Versioned, P2P File System},
	howpublished = {arXiv preprint arXiv:1407.3561},
	year         = {2014}
}

@article{pan2010survey,
	author  = {Sinno Jialin Pan and Qiang Yang},
	title   = {A Survey on Transfer Learning},
	journal = {IEEE Transactions on Knowledge and Data Engineering},
	volume  = {22},
	number  = {10},
	pages   = {1345--1359},
	year    = {2010}
}

@inproceedings{bhatia2022dfl,
	author    = {Laveen Bhatia and Saeed Samet},
	title     = {Decentralized Federated Learning: A Comprehensive Survey and a New Blockchain-based Data Evaluation Scheme},
	booktitle = {2022 Fourth International Conference on Blockchain Computing and Applications (BCCA)},
	pages     = {289--296},
	year      = {2022},
	doi       = {10.1109/BCCA55292.2022.9922390}
}

@article{alizouaghi2025ipfs,
	author  = {Yousra Ali Zouaghi and Meriama Mahamdioua and Atidel Lahoulou and Seloua Chettibi},
	title   = {Privacy preserving biometric authentication based on fully homomorphic encryption, blockchain, and IPFS data storage},
	journal = {Multimedia Tools and Applications},
	volume  = {84},
	number  = {34},
	pages   = {42293--42319},
	year    = {2025},
	doi     = {10.1007/s11042-025-20817-y}
}

@misc{bib:secp,
    author = {{Certicom Research}},
    title = {Standards for Efficient Cryptography - {SEC 2}: Recommended Elliptic Curve Domain Parameters},
    howpublished = {https://www.secg.org/sec2-v2.pdf}
}

@article{haakegaard2015elliptic,
  title={The elliptic curve diffie-hellman (ecdh)},
  author={Haakegaard, Rakel and Lang, Joanna},
  journal={Online at https://koclab. cs. ucsb. edu/teaching/ecc/project/2015Projects/Haakegaard+ Lang. pdf},
  volume={5},
  pages={74},
  year={2015}
}

\end{document}